\documentclass[aps,prl,twocolumn,amsmath,amssymb,showpacs,footnote,superscriptaddress,reprint]{revtex4-1}

\usepackage{graphicx}
\usepackage{dcolumn}
\usepackage{bm}

\usepackage[dvipdfm]{hyperref}

\hypersetup
    {
        colorlinks,%
        citecolor=blue,%
        linkcolor=blue,%
        urlcolor=blue,%
    }

\begin{document}

\title{Giant black hole ringings induced by massive gravity}

\author{Yves D\'ecanini}
\email{decanini@univ-corse.fr}

\affiliation{Equipe Physique
Th\'eorique - Projet COMPA, \\ SPE, UMR 6134 du CNRS
et de l'Universit\'e de Corse,\\
Universit\'e de Corse, BP 52, F-20250 Corte,
France}

\author{Antoine Folacci}
\email{folacci@univ-corse.fr}

\affiliation{Equipe Physique
Th\'eorique - Projet COMPA, \\ SPE, UMR 6134 du CNRS
et de l'Universit\'e de Corse,\\
Universit\'e de Corse, BP 52, F-20250 Corte,
France}

\author{Mohamed Ould El Hadj}
\email{ould-el-hadj@univ-corse.fr}

\affiliation{Equipe Physique
Th\'eorique - Projet COMPA, \\ SPE, UMR 6134 du CNRS
et de l'Universit\'e de Corse,\\
Universit\'e de Corse, BP 52, F-20250 Corte,
France}

\date{\today}

\begin{abstract}

A distorted black hole radiates gravitational waves in order to settle down in one of the geometries permitted by the no-hair theorem. During that relaxation phase, a characteristic damped ringing is generated. It can be theoretically constructed from the black hole quasinormal frequencies (which govern its oscillating behavior and its decay) and from the associated excitation factors (which determine intrinsically its amplitude) by carefully taking into account the source of the distortion. Here, by considering the Schwarzschild black hole in the framework of massive gravity, we show that the excitation factors have an unexpected strong resonant behavior leading to giant ringings which are, moreover, slowly decaying. Such extraordinary black hole ringings could be observed by the next generations of gravitational wave detectors and allow us to test the various massive gravity theories or their absence could be used to impose strong constraints on the graviton mass.

\end{abstract}


\maketitle

{\it Introduction.}--- Gravitational waves, a major prediction of Einstein's general relativity,
should be observed directly in a near future by the next generations of gravitational wave detectors. Another fascinating prediction of Einstein's theory, the existence of black holes (BHs), should be simultaneously confirmed. Indeed, if in its final stage the astrophysical process generating the observed gravitational radiation involves a distorted BH, the signal is then dominated, at intermediate time scales, by a characteristic damped ringing. It is due to the BH which radiates away all its distortions in the form of gravitational waves and relaxes toward a state permitted by the no-hair theorem. The frequencies and decay rates describing that ringing define the complex resonance spectrum of the BH. They are linked to its quasinormal modes (QNMs) \cite{Nollert:1999ji,Kokkotas:1999bd,Berti:2009kk,Konoplya:2011qq}, i.e., to those mode solutions of the wave equation which propagate inward at the horizon and outward at spatial infinity, and they can be considered as the BH fingerprint : they indicate beyond all doubt the existence of a horizon and they could be used to determine unambiguously the mass as well as the angular momentum of the BH.

These last years, generalizations of general relativity mediated by a massive spin-2 particle are the subject of intense activity (see Ref.~\cite{Hinterbichler:2011tt} for a recent review). They have their source in the 70-year old Fierz-Pauli theory \cite{Fierz:1939ix}. They are motivated by purely theoretical considerations (the study of the deformations of general relativity with the graviton mass as deformation parameter) but they also arise from field theories in spacetimes with extra dimensions. Furthermore, and this is surely the main raison of their success, they could explain, without dark energy, the accelerated expansion of the present Universe. Of course, a hypothetical massive graviton is surely an ultralight particle (see Ref.~\cite{Goldhaber:2008xy} for a description of the experimental constraints on the graviton mass but note that the mass limit strongly depends on the theory considered).

It is therefore natural to study BH perturbations and to reconsider gravitational radiation from BHs in massive gravity. It is only very recently that some works on this subject by Babichev and Fabbri \cite{Babichev:2013una} and by Brito, Cardoso and Pani \cite{Brito:2013wya,Brito:2013yxa} have been achieved. They mainly discuss the fundamental problem of BH stability and therefore the existence of BHs in massive gravity. Of course, this depends on the model of gravity considered but, once we assume stability, it becomes really interesting to work on the structure of the signal emitted by a distorted BH with in mind the possibility to test, in a near future, the various massive spin-2 field theories with gravitational wave detectors.

Since the seventies, an increasing number of frequency- and time-domain studies dealing with massive fields propagating in BH spacetimes have highlighted the important modifications induced by the mass parameter which concern more or less directly the signal emitted by a distorted BH : (i) the resonance spectrum is enriched by the complex frequencies corresponding to quasibound states (see Refs.~\cite{Deruelle:1974zy,Damour:1976kh,Zouros:1979iw,Detweiler:1980uk} for important pioneering works and Ref.~\cite{Brito:2013wya} for a recent study in massive gravity); (ii) as the mass parameter increases, the quasinormal frequencies migrate in the complex plane, a behavior observed numerically by various authors (see,
e.g., Refs.~\cite{Simone:1991wn,Konoplya:2004wg} for pioneering works and Ref.~\cite{Brito:2013wya} for a recent study in massive gravity) and analytically described recently \cite{Hod:2011zzd,Decanini:2011eh}; (iii) at very late time (i.e., after the quasinormal ringing), the signal emitted by the relaxing BH is not described by the usual power-law tail behavior \cite{Price:1971fb} but, roughly speaking, by oscillations with a slowly-decaying amplitude (see, e.g., Ref.~\cite{Burko:2004jn} as well as Ref.~\cite{Hod:2013dka} for a recent study in massive gravity).

In a future article \cite{DFOEH1}, we intend to discuss as fully as possible a new effect with amazing consequences : for massive bosonic fields in the Schwarzschild spacetime, the excitation factors of the QNMs have a strong resonant behavior which induces giant ringings. It is a totally unexpected effect. Indeed, until now it was assumed that, for massive fields, quasinormal ringings are less easily excited (see the introduction of Ref.~\cite{Rosa:2011my} and references therein). We shall describe this effect numerically and confirm it analytically from semiclassical considerations based on the properties of the unstable circular geodesics on which a massive particle can orbit the BH.

In this letter, we only consider the massive spin-2 field because of its theoretical importance and due to the fascinating observational consequences that our results predict for such a field. We limit our study to the Fierz-Pauli theory in the Schwarzschild spacetime \cite{Brito:2013wya} which can be obtained, e.g., by linearization of the ghost-free bimetric theory of Hassan, Schmidt-May and von Strauss discussed in Ref.~\cite{Hassan:2012wr} and which is inspired by the fundamental work of de Rham, Gabadadze and Tolley \cite{deRham:2010ik,deRham:2010kj}. Furthermore, we mainly focus on the odd-parity $\ell=1$ mode of this field (similar results can be obtained for the other modes - see also Ref.~\cite{DFOEH1}) and on the associated QNMs. Our letter is organized as follows. We first establish numerically the resonant behavior of the excitation factor of the $(\ell=1,n=0)$ QNM (and briefly discuss its overtones). It occurs in a large domain around a critical value of the mass parameter where the QNM is, in addition, weakly damped. As a consequence, it induces giant and slowly decaying ringings. These are constructed directly from the retarded Green function (an intrinsic point of view) and we compare them with the ringing generated by the odd-parity $(\ell=2,n=0)$ QNM of the massless theory which, in the context of Einstein gravity, provides one of the most important contribution to the BH ringing. With in mind observational considerations, it is necessary to check that a realistic perturbation describing the BH distortion does not neutralize the resonant effect previously discussed and to study ringings constructed from the quasinormal excitation coefficients (an extrinsic point of view) because they permit us to include the contribution of the perturbation into the BH response \cite{Berti:2006wq}. We then describe the perturbation by an initial value problem, an approach which has regularly provided interesting results \cite{Leaver:1986gd,Andersson:1996cm,Berti:2006wq}. We show that the excitation coefficient of the $(\ell=1,n=0)$ QNM also has a resonant behavior which still leads to giant and slowly decaying ringings. In a conclusion, we discuss some possible extensions of our work and its interest for gravitational wave astrophysics and theoretical physics.

Throughout this letter, we display our numerical results by using the dimensionless coupling constant ${\tilde \alpha}=2M\mu /{m_\mathrm{P}}^2$ (here $M$, $\mu$ and $m_\mathrm{P}= \sqrt{\hbar c /G} $ denote respectively the mass of the BH, the rest mass of the graviton and the Planck mass). We adopt units such that $\hbar = c = G = 1$ and assume a harmonic time dependence $\exp(-i\omega t)$ for the spin-2 field. We describe the exterior of the Schwarzschild BH by using both the radial coordinate $r \in ]2M,+\infty[$ and the so-called tortoise coordinate $r_\ast \in ]-\infty,+\infty[$ given by $r_\ast(r)=r+2M \ln[r/(2M)-1]$.

{\it Resonant behavior of the quasinormal excitation factors and associated ``intrinsic" giant ringings.}--- In Schwarzschild spacetime, the partial amplitude $\phi (t,r)$ describing the odd-parity $\ell=1$ mode of the massive spin-2 field satisfies \cite{Brito:2013wya} (the angular momentum index $\ell=1 $ will be, from now on, suppressed in all the formulas)
\begin{equation}\label{Phi_ell1}
\left[-\frac{\partial^2 }{\partial t^2}+\frac{\partial^2}{\partial r_\ast^2}-V(r)  \right] \phi (t,r)=0
\end{equation}
with the effective potential $V(r)$ given by
\begin{equation}\label{pot_RW_Schw}
V(r) = \left(1-\frac{2M}{r} \right) \left(\mu^2+
\frac{6}{r^2} -\frac{16M}{r^3}\right).
\end{equation}
The associated retarded Green function can be written as
\begin{equation}\label{Gret_om}
G_\mathrm{ret}(t;r,r')=-\int_{-\infty +ic}^{+\infty +ic}  \frac{d\omega}{2\pi}  \frac{\phi^\mathrm{in}_{\omega}(r_<) \phi^\mathrm{up}_{\omega}(r_>)}{W (\omega)} e^{-i\omega t}
\end{equation}
\noindent where $c>0$, $r_< =\mathrm{min} (r,r')$, $r_> =\mathrm{max} (r,r')$ and with $W (\omega)$ denoting the Wronskian of the functions $\phi^\mathrm{in}_{\omega}$ and $\phi^\mathrm{up}_{\omega}$. These two functions are linearly independent solutions of the Regge-Wheeler equation
\begin{equation}\label{RW}
\frac{d^2 \phi_{\omega}}{dr_\ast^2} + \left[ \omega^2 -V(r)\right]  \phi_{\omega}=0.
\end{equation}
When $\mathrm{Im} (\omega) > 0$, $\phi^\mathrm{in}_{\omega}$ is uniquely defined by its ingoing behavior at the event horizon, i.e., for $r_\ast \to -\infty$ $\phi^\mathrm{in}_{\omega} (r) \sim \exp[-i\omega r_\ast]$
and, at spatial infinity, i.e., for $r_\ast \to +\infty$, it has an
asymptotic behavior of the form
\begin{eqnarray}\label{bc_2_in}
& & \phi^\mathrm{in}_{\omega}(r) \sim
 \sqrt{ \frac{\omega}{p(\omega)}} \left[A^{(-)} (\omega) e^{-i[p(\omega)
r_\ast + [M\mu^2/p(\omega)] \ln(r/M)]}\right. \nonumber \\
& & \quad \quad  \left. + A^{(+)} (\omega) e^{+i[p(\omega) r_\ast +
[M\mu^2/p(\omega)] \ln(r/M)]} \right].
\end{eqnarray}
Similarly, $\phi^\mathrm{up}_{\omega}$ is uniquely defined by its outgoing behavior at spatial infinity, i.e., for $r_\ast \to +\infty$, $\phi^\mathrm{up}_{\omega} (r) \sim  \sqrt{ \omega/p(\omega)} \exp \lbrace{+i[p(\omega) r_\ast +
[M\mu^2/p(\omega)] \ln(r/M)] \rbrace}$ and, at the horizon, i.e., for $r_\ast \to -\infty$ it has an asymptotic behavior of the form
\begin{equation}\label{bc_2_up}
\phi^\mathrm{up}_{\omega}(r) \sim
B^{(-)} (\omega) e^{-i\omega r_\ast}  + B^{(+)} (\omega) e^{+i\omega r_\ast}.
\end{equation}
Here $p(\omega)=\sqrt{\omega^2 - \mu^2 }$ denotes the ``wave number"
while $A^{(-)} (\omega)$, $A^{(+)} (\omega)$, $B^{(-)} (\omega)$ and $B^{(+)} (\omega)$ are complex amplitudes which, like the $\mathrm{in}$- and $\mathrm{up}$- modes, can be defined by analytic continuation in the full complex $\omega$-plane (or, more precisely, in a well-chosen multi-sheeted Riemann surface). By evaluating the Wronskian $W (\omega)$ at $r_\ast \to -\infty$ and $r_\ast \to +\infty$, we obtain $W (\omega) =2i\omega A^{(-)} (\omega) = 2i\omega B^{(+)} (\omega)$.

If the Wronskian $W (\omega)$ vanishes, the functions $\phi^\mathrm{in}_{\omega}$ and $\phi^\mathrm{up}_{\omega}$ are linearly dependent and propagate inward at the horizon and outward at spatial infinity, a behavior which defines the QNMs. The zeros of the Wronskian lying in the lower part of the complex $\omega$-plane are the frequencies of the $\ell=1$ QNMs. They are symmetrically distributed with respect to the imaginary $\omega$-axis. The contour of integration in Eq.~(\ref{Gret_om}) may be deformed in order to capture them \cite{Leaver:1986gd}. By Cauchy's Theorem, we can extract from the retarded Green function (\ref{Gret_om}) a residue series over the quasinormal frequencies $\omega_n$ lying in the fourth quadrant of the complex $\omega$-plane. We then obtain the contribution describing the BH ringing. It is given by
\begin{eqnarray}\label{Gret_ell_QNM}
&& G_\mathrm{ret}^\mathrm{QNM}(t;r,r')=2 \, \mathrm{Re} \left[ \sum_n {\cal B}_{n}
{\tilde \phi}_{\omega_n}(r) {\tilde \phi}_{\omega_n}(r') \right. \nonumber\\
&&  \left. \vphantom{\sum_n} \times e^{-i\omega_n t + ip(\omega_n)r_\ast + ip(\omega_n)r'_\ast +
i[M\mu^2/p(\omega_n)] \ln(rr'/M^2)} \right]
\end{eqnarray}
where
\begin{equation}\label{Excitation F}
{\cal B}_{n} = \left(\frac{1}{2 p(\omega)} \frac{A^{(+)} (\omega)}{\frac{dA^{(-)} (\omega)}{d\omega}}   \right)_{\omega=\omega_n}
\end{equation}
denotes the excitation factor corresponding to the complex frequency $\omega_n$. In Eq.~(\ref{Gret_ell_QNM}), the modes ${\tilde \phi}_{\omega_n}(r)$ are defined by normalizing the modes $\phi^\mathrm{in}_{\omega_n}(r)$ so that ${\tilde \phi}_{\omega_n}(r) \sim 1$ as $r \to +\infty $. In the sum, $n=0$ corresponds to the fundamental QNM (i.e., the least damped one) and $n=1,2,\dots $ to the overtones.

Quasinormal retarded Green functions such as (\ref{Gret_ell_QNM}) do not provide physically relevant results at ``early times" due to their exponentially divergent behavior as $t$ decreases. It is necessary to determine, from physical considerations, the time beyond which they can be used and such a time is the starting time $t_\mathrm{start}$ of the BH ringing. $t_\mathrm{start}$ can be ``easily" obtained for massless fields (see, e.g., Ref.~\cite{Berti:2006wq}). Indeed, we first note that the QNMs are semiclassically associated with the peak of the effective potential located close to $r_\ast \approx 0 $. Then, by assuming that the source at $r'_\ast$ and the observer at $r_\ast$ are far from the BH (i.e., that $r_\ast,r'_\ast \gg 2M$) we have $t_\mathrm{start} \approx r_\ast + r'_\ast$ (it is approximatively the time taken for the signal to travel from the source to the peak of the potential and then to reach the observer). For massive fields, the previous considerations must be slightly modified. We take into account the dispersive behavior of the QNMs and define $t_\mathrm{start}$ from group velocities. From the dispersion relation $p(\omega)=\sqrt{\omega^2 - \mu^2 }$, we can show that the group velocity corresponding to the quasinormal frequency $\omega_n$ is given by $v_\mathrm{g}=\mathrm{Re}[p(\omega_n)]/\mathrm{Re}[\omega_n]$. Because the peak of the effective potential still remains located close to $r_\ast \approx 0 $, we then obtain $t_\mathrm{start} \approx (r_\ast + r'_\ast)\mathrm{Re}[\omega_n]/\mathrm{Re}[p(\omega_n)]$ (here, we neglect the contribution of $[M\mu^2/p(\omega_n)] \ln(rr'/M^2)$).

\begin{figure}
\includegraphics[height=2cm,width=8.5cm]{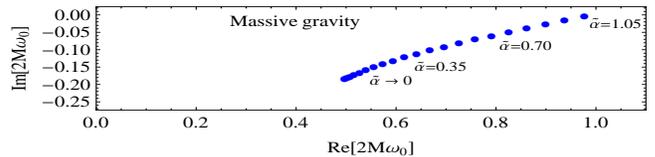}
\setlength\abovecaptionskip{0.25ex}
\vspace{-0.05cm}
\caption{\label{fig:OM_n=0} Complex frequency $\omega_0$ of the odd-parity $(\ell=1,n=0)$ QNM.}
\end{figure}

\begin{figure}
\includegraphics[height=3.3cm,width=8.5cm]{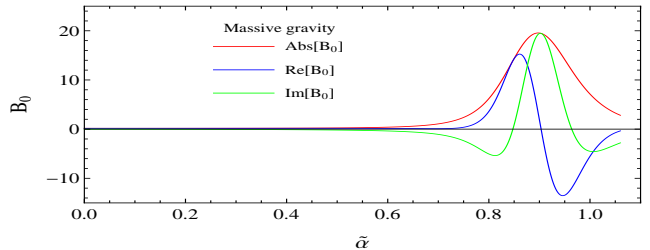}
\setlength\abovecaptionskip{0.25ex}
\vspace{-0.07cm}
\caption{\label{fig:B0} Resonant behavior of the excitation factor ${\cal B}_0$ of the odd-parity $(\ell=1,n=0)$ QNM. The maximum of $|{\cal B}_0|$ occurs for the critical value ${\tilde \alpha} \approx 0.89757$; we then have $2M\omega_0 \approx 0.85969073 - 0.03878222 i$ and ${\cal B}_0 \approx 3.25237 + 19.28190 i$.}
\end{figure}

In Fig.~\ref{fig:OM_n=0}, we display the effect of the graviton mass on $\omega_0$ (see also Fig.~2 in Ref.~\cite{Brito:2013wya}) and in Fig.~\ref{fig:B0}, we exhibit the strong resonant behavior of ${\cal B}_0$ occurring around the critical value ${\tilde \alpha} \approx 0.90$. The same kind of resonant behavior exists for the excitation factors ${\cal B}_n $ with $n \not= 0$ but the resonance amplitude decreases rapidly as the overtone index $n$ increases.


\begin{figure}
\includegraphics[height=3.3cm,,width=8.5cm]{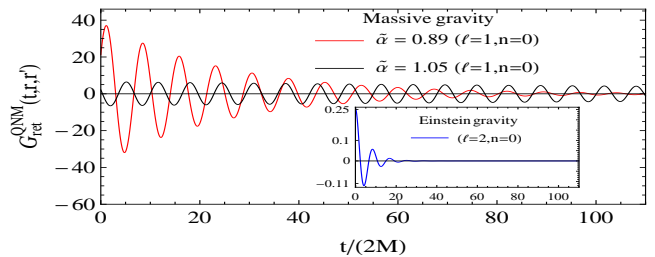}
\setlength\abovecaptionskip{0.25ex}
\vspace{-0.05cm}
\caption{\label{fig:IntrinsicRingings} Extraordinary ``intrinsic" ringings induced by massive gravity and comparison with the ringing generated in Einstein gravity. The results are obtained from (\ref{Gret_ell_QNM}) with $r = 50M$ and $r' = 10M$.}
\end{figure}

The resonant behavior of the excitation factor ${\cal B}_0$ occurring for masses in a range where the QNM is a long-lived mode (see Figs.~\ref{fig:OM_n=0} and \ref{fig:B0}) induces giant ringings which are, moreover, slowly or even very slowly decaying. In Fig.~\ref{fig:IntrinsicRingings}, we plot, for two values of the graviton mass, the BH ``intrinsic" ringings constructed from the quasinormal retarded Green function (\ref{Gret_ell_QNM}) and we compare them to the ringing generated by the odd-parity $(\ell=2,n=0)$ QNM of the massless spin-2 field with quasinormal frequency $2M \omega_{20} \approx 0.74734337 - 0.17792463 i$ and with excitation factor ${\cal B}_{20} \approx 0.12690 + 0.02032 i$. Similar results can be obtained for various locations of the source and the observer. Giant BH ringings also exist for $n \not= 0$ but with less impressive characteristics.

{\it Resonant behavior of the quasinormal excitation coefficients and associated ``extrinsic" giant ringings.}--- In the previous section, we focussed on ``intrinsic" ringings, i.e., on ringings directly constructed from the quasinormal retarded Green function and therefore depending only on the BH properties. Of course, with in mind astrophysical applications, it is now necessary to check that giant ringings also exist in the presence of a realistic perturbation or, in other words, that the convolution of the source of the perturbation with the retarded Green function does not modify, in a fundamental way, our results. This is a complex problem and, in this letter, we just discuss some of its elementary aspects.

Some years ago, dealing with the observability of quasinormal ringings, Andersson and Glampedakis associated to each QNM an effective amplitude $h_\mathrm{eff}$ achievable after matched filtering \cite{Andersson:1999wj,Glampedakis:2001js}. For the massive QNMs considered here, it reads
\begin{equation}\label{Eff_Amp}
h_\mathrm{eff} \sim   \mathrm{Re} \left[2 \sqrt{-\mathrm{Re}(\omega_n)/\mathrm{Im}(\omega_n)} \, p(\omega_n) {\cal B}_n \right].
\end{equation}
We can use it in order to determine the effective amplitude of the ringings generated by the odd-parity $(\ell=1,n=0)$ QNM of massive gravity; we obtain $|h_\mathrm{eff}| \approx 54.74$ for ${\tilde \alpha}=0.89$ and $|h_\mathrm{eff}| \approx 13.02$ for ${\tilde \alpha}=1.05$. For the ringing generated by the odd-parity $(\ell=2,n=0)$ QNM of Einstein gravity we have $|h_\mathrm{eff}| \approx 0.40$. This is in perfect agreement with our previous results and suggests that giant BH ringings are astrophysically relevant and observable. However, it should be noted that Andersson-Glampedakis formula must be taken with a pinch of salt. As noted in Ref.~\cite{Berti:2006wq}, it seems helpful only if the quasinormal ringing is excited by localized initial data. So, it is necessary to consider a more general approach.

We now describe the BH perturbation by an initial value problem with Gaussian initial data. More precisely, we consider that, at $t=0$, the partial amplitude $\phi (t,r)$ governed by (\ref{Phi_ell1}) satisfies
\begin{equation}\label{Cauchy_data}
\phi (t=0,r)=\phi_0(r) \equiv \phi_0 \exp \left[-\frac{a^2}{(2M)^2} (r_\ast - \beta)^2 \right]
\end{equation}
and $\partial_t\phi (t=0,r)=0$. By Green's Theorem, we can show that the time evolution of $\phi (t,r)$ is described, for $t>0$, by $\phi (t,r)=\int \partial_t G_\mathrm{ret}(t;r,r') \phi_0(r')   dr'_\ast.$ We can insert (\ref{Gret_om}) into this expression and deform again the contour of integration on $\omega$ in order to capture the contributions of the QNMs. We then isolate the BH ringing generated by the initial data:
\begin{eqnarray}\label{TimeEvolution_QNM}
&&\phi^\mathrm{QNM} (t,r)= 2 \, \mathrm{Re} \left[ \sum_n i\omega_n {\cal C}_n \right. \nonumber\\
&&  \left. \vphantom{\sum_n} \qquad \times e^{-i\omega_n t+ip(\omega_n)r_\ast +
i[M\mu^2/p(\omega_n)] \ln(r/M)} \right].
\end{eqnarray}
Here ${\cal C}_n$ denotes the excitation coefficient of the QNM with overtone index $n$. It takes explicitly into account the role of the BH perturbation and is given by
 \begin{equation}\label{EC}
{\cal C}_n={\cal B}_n \int \frac{\phi_0(r')\phi^\mathrm{in}_{\omega_n}(r')}{\sqrt{\omega_n/p(\omega_n)}A^{(+)}(\omega_n)}   dr'_\ast.
\end{equation}

\begin{figure}
\includegraphics[height=3.3cm,,width=8.5cm]{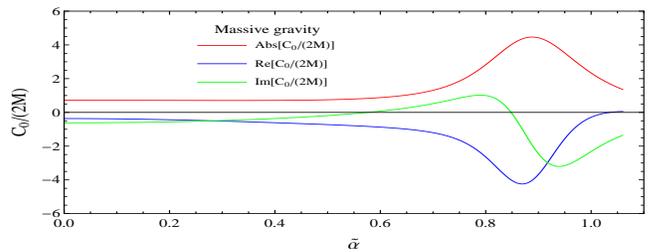}
\setlength\abovecaptionskip{0.25ex}
\vspace{-0.05cm}
\caption{\label{fig:C0} Resonant behavior, in massive gravity, of the excitation coefficient ${\cal C}_0$ of the odd-parity $(\ell=1,n=0)$ QNM. It is obtained from (\ref{EC}) by using (\ref{Cauchy_data}) with $\phi_0=1$, $a=1$ and $\beta=10M$. The maximum of $|{\cal C}_0|/(2M)$ occurs for the critical value ${\tilde \alpha} \approx 0.88808$; we then have $2M\omega_0 \approx 0.85277076 - 0.04084908 i$ and ${\cal C}_0/(2M) \approx -4.02613 - 1.93037 i$.}
\end{figure}

The excitation coefficients ${\cal C}_n$, like the excitation factors ${\cal B}_n$, have a resonant behavior but it is now more attenuated. Moreover, the maximum amplitude of the resonance is slightly shifted but still occurs for masses in a range where the QNM is a long-lived mode. In Fig.~\ref{fig:C0}, we exhibit the strong resonant behavior of ${\cal C}_0$ for particular values of the parameters defining the initial data (\ref{Cauchy_data}). It occurs around the critical value ${\tilde \alpha} \approx 0.89$ and is rather similar to the behavior of the corresponding excitation factor ${\cal B}_0$. It should be noted that we have checked that it depends very little on the parameters defining the initial data (\ref{Cauchy_data}) (see Ref.~\cite{DFOEH1} for a detailed study). Of course, for overtones, the resonance is more and more attenuated as the overtone index $n$ increases so, the ringings generated by the fundamental QNM are certainly the most interesting.

\begin{figure}
\includegraphics[height=3.3cm,,width=8.5cm]{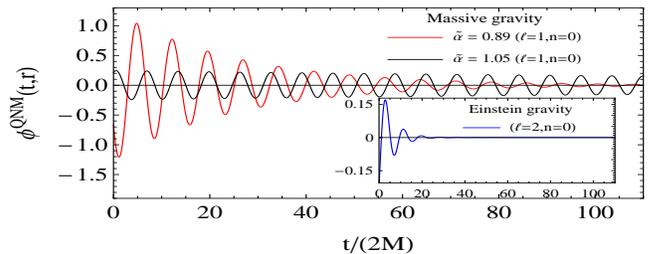}
\setlength\abovecaptionskip{0.25ex}
\vspace{-0.05cm}
\caption{\label{fig:ExtrinsicRingings} Extraordinary ``extrinsic" ringings induced by massive gravity and comparison with the ringing generated in Einstein gravity. The results are obtained from (\ref{TimeEvolution_QNM}) with $r = 50M$ by using (\ref{Cauchy_data}) with $\phi_0=1$, $a=1$ and $\beta=10M$.}
\end{figure}

In Fig.~\ref{fig:ExtrinsicRingings}, we plot, for the two values of the graviton mass considered in Fig.~\ref{fig:IntrinsicRingings}, the BH ``extrinsic" ringings defined by (\ref{TimeEvolution_QNM}) and we compare them to the ringing generated by the odd-parity $(\ell=2,n=0)$ QNM of the massless spin-2 field. This last one is constructed by noting that, for the same initial data, the excitation coefficient of the QNM is given by ${\cal C}_{20}/(2M) \approx 0.50761 - 0.29210 i$. Similar results as those displayed in Fig.~\ref{fig:ExtrinsicRingings} can be obtained for various values of the parameters defining the initial data (\ref{Cauchy_data}) and for various locations of the observer. Even if the role of the perturbation is taken into account, extraordinary BH ringings exist.

{\it Conclusion.}--- In this letter, by considering the massive spin-2 field in Schwarzschild spacetime, we have pointed out a new effect in BH physics : the existence around particular values of the mass parameter of a strong resonant behavior for the excitation factors of the QNMs with, as a consequence, the existence of giant and slowly-decaying ringings. Such results are, in fact, a general feature of massive field theories in the Schwarzschild BH \cite{DFOEH1}. It would be interesting to study more realistic perturbations than the distortion described here by an initial value problem (e.g., the excitation of the BH by a particle falling radially or plunging), to consider alternative massive gravity theories and to extend our study to the Kerr BH. Finally, we would like to note that :

\qquad (i) The Schwarzschild BH interacting with a massive spin-2 field is in general unstable \cite{Babichev:2013una,Brito:2013wya} (see, however, Ref.~\cite{Brito:2013yxa}). In the context of the theory considered here, the instability is due to the behavior of the propagating $\ell=0$ mode \cite{Brito:2013wya}. It is a ``low-mass" instability which disappears for ${\tilde \alpha}$ above the threshold value ${\tilde \alpha}_c \approx 0.86$. So, the giant ringings predicted here occurring near and above the critical value ${\tilde \alpha} \approx 0.89$ are physically relevant. 

\qquad (ii) Even if the graviton is an ultralight particle, when it interacts with a supermassive BH, values of the coupling constant ${\tilde \alpha}$ leading to giant ringings can be easily reached. Indeed, supermassive BHs have their masses lying approximately between $10^6\, \mathrm{M}_{\odot}$ and $2\times 10^{10}\, \mathrm{M}_{\odot}$; so, if we assume that $\mu \approx 1.35\times 10^{-55} \, \mathrm{kg}$ (it seems to be the superior limit of the graviton mass in the framework of the ordinary Fierz-Pauli theory \cite{Finn:2001qi}), we have ${\tilde \alpha}$ lying approximately between $10^{-3}$ and $20$. As a consequence, due to the enormous number of supermassive BHs in the Universe, the extraordinary BH ringings discussed here could be observed by the next generations of gravitational wave detectors and used to test the various massive gravity theories or their absence could allow us to impose strong constraints on the graviton mass and to support, in a new way, Einstein's general relativity.

{\it Acknowledgments.}--- We wish to thank Andrei Belokogne for discussions and the ``Collectivit\'e Territoriale de Corse" for its support through the COMPA project.

\bibliography{Giant_BH_ringing}

\begin{thebibliography}{32}%
\makeatletter
\providecommand \@ifxundefined [1]{%
 \@ifx{#1\undefined}
}%
\providecommand \@ifnum [1]{%
 \ifnum #1\expandafter \@firstoftwo
 \else \expandafter \@secondoftwo
 \fi
}%
\providecommand \@ifx [1]{%
 \ifx #1\expandafter \@firstoftwo
 \else \expandafter \@secondoftwo
 \fi
}%
\providecommand \natexlab [1]{#1}%
\providecommand \enquote  [1]{``#1''}%
\providecommand \bibnamefont  [1]{#1}%
\providecommand \bibfnamefont [1]{#1}%
\providecommand \citenamefont [1]{#1}%
\providecommand \href@noop [0]{\@secondoftwo}%
\providecommand \href [0]{\begingroup \@sanitize@url \@href}%
\providecommand \@href[1]{\@@startlink{#1}\@@href}%
\providecommand \@@href[1]{\endgroup#1\@@endlink}%
\providecommand \@sanitize@url [0]{\catcode `\\12\catcode `\$12\catcode
  `\&12\catcode `\#12\catcode `\^12\catcode `\_12\catcode `\%12\relax}%
\providecommand \@@startlink[1]{}%
\providecommand \@@endlink[0]{}%
\providecommand \url  [0]{\begingroup\@sanitize@url \@url }%
\providecommand \@url [1]{\endgroup\@href {#1}{\urlprefix }}%
\providecommand \urlprefix  [0]{URL }%
\providecommand \Eprint [0]{\href }%
\providecommand \doibase [0]{http://dx.doi.org/}%
\providecommand \selectlanguage [0]{\@gobble}%
\providecommand \bibinfo  [0]{\@secondoftwo}%
\providecommand \bibfield  [0]{\@secondoftwo}%
\providecommand \translation [1]{[#1]}%
\providecommand \BibitemOpen [0]{}%
\providecommand \bibitemStop [0]{}%
\providecommand \bibitemNoStop [0]{.\EOS\space}%
\providecommand \EOS [0]{\spacefactor3000\relax}%
\providecommand \BibitemShut  [1]{\csname bibitem#1\endcsname}%
\let\auto@bib@innerbib\@empty
\bibitem [{\citenamefont {Nollert}(1999)}]{Nollert:1999ji}%
  \BibitemOpen
  \bibfield  {author} {\bibinfo {author} {\bibfnamefont {H.-P.}\ \bibnamefont
  {Nollert}},\ }\href {\doibase 10.1088/0264-9381/16/12/201} {\bibfield
  {journal} {\bibinfo  {journal} {Classical Quantum Gravity}\ }\textbf
  {\bibinfo {volume} {16}},\ \bibinfo {pages} {R159} (\bibinfo {year}
  {1999})}\BibitemShut {NoStop}%
\bibitem [{\citenamefont {Kokkotas}\ and\ \citenamefont
  {Schmidt}(1999)}]{Kokkotas:1999bd}%
  \BibitemOpen
  \bibfield  {author} {\bibinfo {author} {\bibfnamefont {K.~D.}\ \bibnamefont
  {Kokkotas}}\ and\ \bibinfo {author} {\bibfnamefont {B.~G.}\ \bibnamefont
  {Schmidt}},\ }\href {\doibase 10.12942/lrr-1999-2} {\bibfield  {journal}
  {\bibinfo  {journal} {Living Rev.\ Rel.}\ }\textbf {\bibinfo {volume} {2}},\
  \bibinfo {pages} {2} (\bibinfo {year} {1999})},\ \Eprint
  {http://arxiv.org/abs/gr-qc/9909058} {arXiv:gr-qc/9909058} \BibitemShut
  {NoStop}%
\bibitem [{\citenamefont {Berti}\ \emph {et~al.}(2009)\citenamefont {Berti},
  \citenamefont {Cardoso},\ and\ \citenamefont {Starinets}}]{Berti:2009kk}%
  \BibitemOpen
  \bibfield  {author} {\bibinfo {author} {\bibfnamefont {E.}~\bibnamefont
  {Berti}}, \bibinfo {author} {\bibfnamefont {V.}~\bibnamefont {Cardoso}}, \
  and\ \bibinfo {author} {\bibfnamefont {A.~O.}\ \bibnamefont {Starinets}},\
  }\href {\doibase 10.1088/0264-9381/26/16/163001} {\bibfield  {journal}
  {\bibinfo  {journal} {Classical Quantum Gravity}\ }\textbf {\bibinfo {volume}
  {26}},\ \bibinfo {pages} {163001} (\bibinfo {year} {2009})},\ \Eprint
  {http://arxiv.org/abs/0905.2975} {arXiv:0905.2975 [gr-qc]} \BibitemShut
  {NoStop}%
\bibitem [{\citenamefont {Konoplya}\ and\ \citenamefont
  {Zhidenko}(2011)}]{Konoplya:2011qq}%
  \BibitemOpen
  \bibfield  {author} {\bibinfo {author} {\bibfnamefont {R.}~\bibnamefont
  {Konoplya}}\ and\ \bibinfo {author} {\bibfnamefont {A.}~\bibnamefont
  {Zhidenko}},\ }\href {\doibase 10.1103/RevModPhys.83.793} {\bibfield
  {journal} {\bibinfo  {journal} {Rev.\ Mod.\ Phys.}\ }\textbf {\bibinfo
  {volume} {83}},\ \bibinfo {pages} {793} (\bibinfo {year} {2011})},\ \Eprint
  {http://arxiv.org/abs/1102.4014} {arXiv:1102.4014 [gr-qc]} \BibitemShut
  {NoStop}%
\bibitem [{\citenamefont {Hinterbichler}(2012)}]{Hinterbichler:2011tt}%
  \BibitemOpen
  \bibfield  {author} {\bibinfo {author} {\bibfnamefont {K.}~\bibnamefont
  {Hinterbichler}},\ }\href {\doibase 10.1103/RevModPhys.84.671} {\bibfield
  {journal} {\bibinfo  {journal} {Rev.\ Mod.\ Phys.}\ }\textbf {\bibinfo
  {volume} {84}},\ \bibinfo {pages} {671} (\bibinfo {year} {2012})},\ \Eprint
  {http://arxiv.org/abs/1105.3735} {arXiv:1105.3735 [hep-th]} \BibitemShut
  {NoStop}%
\bibitem [{\citenamefont {Fierz}\ and\ \citenamefont
  {Pauli}(1939)}]{Fierz:1939ix}%
  \BibitemOpen
  \bibfield  {author} {\bibinfo {author} {\bibfnamefont {M.}~\bibnamefont
  {Fierz}}\ and\ \bibinfo {author} {\bibfnamefont {W.}~\bibnamefont {Pauli}},\
  }\href {\doibase 10.1098/rspa.1939.0140} {\bibfield  {journal} {\bibinfo
  {journal} {Proc.\ Roy.\ Soc.\ Lond.\ A}\ }\textbf {\bibinfo {volume} {173}},\
  \bibinfo {pages} {211} (\bibinfo {year} {1939})}\BibitemShut {NoStop}%
\bibitem [{\citenamefont {Goldhaber}\ and\ \citenamefont
  {Nieto}(2010)}]{Goldhaber:2008xy}%
  \BibitemOpen
  \bibfield  {author} {\bibinfo {author} {\bibfnamefont {A.~S.}\ \bibnamefont
  {Goldhaber}}\ and\ \bibinfo {author} {\bibfnamefont {M.~M.}\ \bibnamefont
  {Nieto}},\ }\href {\doibase 10.1103/RevModPhys.82.939} {\bibfield  {journal}
  {\bibinfo  {journal} {Rev.\ Mod.\ Phys.}\ }\textbf {\bibinfo {volume} {82}},\
  \bibinfo {pages} {939} (\bibinfo {year} {2010})},\ \Eprint
  {http://arxiv.org/abs/0809.1003} {arXiv:0809.1003 [hep-ph]} \BibitemShut
  {NoStop}%
\bibitem [{\citenamefont {Babichev}\ and\ \citenamefont
  {Fabbri}(2013)}]{Babichev:2013una}%
  \BibitemOpen
  \bibfield  {author} {\bibinfo {author} {\bibfnamefont {E.}~\bibnamefont
  {Babichev}}\ and\ \bibinfo {author} {\bibfnamefont {A.}~\bibnamefont
  {Fabbri}},\ }\href {\doibase 10.1088/0264-9381/30/15/152001} {\bibfield
  {journal} {\bibinfo  {journal} {Classical Quantum Gravity}\ }\textbf
  {\bibinfo {volume} {30}},\ \bibinfo {pages} {152001} (\bibinfo {year}
  {2013})},\ \Eprint {http://arxiv.org/abs/1304.5992} {arXiv:1304.5992 [gr-qc]}
  \BibitemShut {NoStop}%
\bibitem [{\citenamefont {Brito}\ \emph
  {et~al.}(2013{\natexlab{a}})\citenamefont {Brito}, \citenamefont {Cardoso},\
  and\ \citenamefont {Pani}}]{Brito:2013wya}%
  \BibitemOpen
  \bibfield  {author} {\bibinfo {author} {\bibfnamefont {R.}~\bibnamefont
  {Brito}}, \bibinfo {author} {\bibfnamefont {V.}~\bibnamefont {Cardoso}}, \
  and\ \bibinfo {author} {\bibfnamefont {P.}~\bibnamefont {Pani}},\ }\href
  {\doibase 10.1103/PhysRevD.88.023514} {\bibfield  {journal} {\bibinfo
  {journal} {Phys.\ Rev.\ D}\ }\textbf {\bibinfo {volume} {88}},\ \bibinfo
  {pages} {023514} (\bibinfo {year} {2013}{\natexlab{a}})},\ \Eprint
  {http://arxiv.org/abs/1304.6725} {arXiv:1304.6725 [gr-qc]} \BibitemShut
  {NoStop}%
\bibitem [{\citenamefont {Brito}\ \emph
  {et~al.}(2013{\natexlab{b}})\citenamefont {Brito}, \citenamefont {Cardoso},\
  and\ \citenamefont {Pani}}]{Brito:2013yxa}%
  \BibitemOpen
  \bibfield  {author} {\bibinfo {author} {\bibfnamefont {R.}~\bibnamefont
  {Brito}}, \bibinfo {author} {\bibfnamefont {V.}~\bibnamefont {Cardoso}}, \
  and\ \bibinfo {author} {\bibfnamefont {P.}~\bibnamefont {Pani}},\ }\href
  {\doibase 10.1103/PhysRevD.87.124024} {\bibfield  {journal} {\bibinfo
  {journal} {Phys.\ Rev.\ D}\ }\textbf {\bibinfo {volume} {87}},\ \bibinfo
  {pages} {124024} (\bibinfo {year} {2013}{\natexlab{b}})},\ \Eprint
  {http://arxiv.org/abs/1306.0908} {arXiv:1306.0908 [gr-qc]} \BibitemShut
  {NoStop}%
\bibitem [{\citenamefont {Deruelle}\ and\ \citenamefont
  {Ruffini}(1974)}]{Deruelle:1974zy}%
  \BibitemOpen
  \bibfield  {author} {\bibinfo {author} {\bibfnamefont {N.}~\bibnamefont
  {Deruelle}}\ and\ \bibinfo {author} {\bibfnamefont {R.}~\bibnamefont
  {Ruffini}},\ }\href {\doibase 10.1016/0370-2693(74)90119-1} {\bibfield
  {journal} {\bibinfo  {journal} {Phys.\ Lett.\ B}\ }\textbf {\bibinfo {volume}
  {52}},\ \bibinfo {pages} {437} (\bibinfo {year} {1974})}\BibitemShut
  {NoStop}%
\bibitem [{\citenamefont {Damour}\ \emph {et~al.}(1976)\citenamefont {Damour},
  \citenamefont {Deruelle},\ and\ \citenamefont {Ruffini}}]{Damour:1976kh}%
  \BibitemOpen
  \bibfield  {author} {\bibinfo {author} {\bibfnamefont {T.}~\bibnamefont
  {Damour}}, \bibinfo {author} {\bibfnamefont {N.}~\bibnamefont {Deruelle}}, \
  and\ \bibinfo {author} {\bibfnamefont {R.}~\bibnamefont {Ruffini}},\ }\href
  {\doibase 10.1007/BF02725534} {\bibfield  {journal} {\bibinfo  {journal}
  {Lett.\ Nuovo Cim.}\ }\textbf {\bibinfo {volume} {15}},\ \bibinfo {pages}
  {257} (\bibinfo {year} {1976})}\BibitemShut {NoStop}%
\bibitem [{\citenamefont {Zouros}\ and\ \citenamefont
  {Eardley}(1979)}]{Zouros:1979iw}%
  \BibitemOpen
  \bibfield  {author} {\bibinfo {author} {\bibfnamefont {T.}~\bibnamefont
  {Zouros}}\ and\ \bibinfo {author} {\bibfnamefont {D.}~\bibnamefont
  {Eardley}},\ }\href {\doibase 10.1016/0003-4916(79)90237-9} {\bibfield
  {journal} {\bibinfo  {journal} {Annals Phys.}\ }\textbf {\bibinfo {volume}
  {118}},\ \bibinfo {pages} {139} (\bibinfo {year} {1979})}\BibitemShut
  {NoStop}%
\bibitem [{\citenamefont {Detweiler}(1980)}]{Detweiler:1980uk}%
  \BibitemOpen
  \bibfield  {author} {\bibinfo {author} {\bibfnamefont {S.~L.}\ \bibnamefont
  {Detweiler}},\ }\href {\doibase 10.1103/PhysRevD.22.2323} {\bibfield
  {journal} {\bibinfo  {journal} {Phys.\ Rev.\ D}\ }\textbf {\bibinfo {volume}
  {22}},\ \bibinfo {pages} {2323} (\bibinfo {year} {1980})}\BibitemShut
  {NoStop}%
\bibitem [{\citenamefont {Simone}\ and\ \citenamefont
  {Will}(1992)}]{Simone:1991wn}%
  \BibitemOpen
  \bibfield  {author} {\bibinfo {author} {\bibfnamefont {L.~E.}\ \bibnamefont
  {Simone}}\ and\ \bibinfo {author} {\bibfnamefont {C.~M.}\ \bibnamefont
  {Will}},\ }\href {\doibase 10.1088/0264-9381/9/4/012} {\bibfield  {journal}
  {\bibinfo  {journal} {Classical Quantum Gravity}\ }\textbf {\bibinfo {volume}
  {9}},\ \bibinfo {pages} {963} (\bibinfo {year} {1992})}\BibitemShut {NoStop}%
\bibitem [{\citenamefont {Konoplya}\ and\ \citenamefont
  {Zhidenko}(2005)}]{Konoplya:2004wg}%
  \BibitemOpen
  \bibfield  {author} {\bibinfo {author} {\bibfnamefont {R.}~\bibnamefont
  {Konoplya}}\ and\ \bibinfo {author} {\bibfnamefont {A.}~\bibnamefont
  {Zhidenko}},\ }\href {\doibase 10.1016/j.physletb.2005.01.078} {\bibfield
  {journal} {\bibinfo  {journal} {Phys.\ Lett.\ B}\ }\textbf {\bibinfo {volume}
  {609}},\ \bibinfo {pages} {377} (\bibinfo {year} {2005})},\ \Eprint
  {http://arxiv.org/abs/gr-qc/0411059} {arXiv:gr-qc/0411059} \BibitemShut
  {NoStop}%
\bibitem [{\citenamefont {Hod}(2011)}]{Hod:2011zzd}%
  \BibitemOpen
  \bibfield  {author} {\bibinfo {author} {\bibfnamefont {S.}~\bibnamefont
  {Hod}},\ }\href {\doibase 10.1103/PhysRevD.84.044046} {\bibfield  {journal}
  {\bibinfo  {journal} {Phys.\ Rev.\ D}\ }\textbf {\bibinfo {volume} {84}},\
  \bibinfo {pages} {044046} (\bibinfo {year} {2011})},\ \Eprint
  {http://arxiv.org/abs/1109.4080} {arXiv:1109.4080 [gr-qc]} \BibitemShut
  {NoStop}%
\bibitem [{\citenamefont {Decanini}\ \emph {et~al.}(2011)\citenamefont
  {Decanini}, \citenamefont {Folacci},\ and\ \citenamefont
  {Raffaelli}}]{Decanini:2011eh}%
  \BibitemOpen
  \bibfield  {author} {\bibinfo {author} {\bibfnamefont {Y.}~\bibnamefont
  {Decanini}}, \bibinfo {author} {\bibfnamefont {A.}~\bibnamefont {Folacci}}, \
  and\ \bibinfo {author} {\bibfnamefont {B.}~\bibnamefont {Raffaelli}},\ }\href
  {\doibase 10.1103/PhysRevD.84.084035} {\bibfield  {journal} {\bibinfo
  {journal} {Phys.\ Rev.\ D}\ }\textbf {\bibinfo {volume} {84}},\ \bibinfo
  {pages} {084035} (\bibinfo {year} {2011})},\ \Eprint
  {http://arxiv.org/abs/1108.5076} {arXiv:1108.5076 [gr-qc]} \BibitemShut
  {NoStop}%
\bibitem [{\citenamefont {Price}(1972)}]{Price:1971fb}%
  \BibitemOpen
  \bibfield  {author} {\bibinfo {author} {\bibfnamefont {R.~H.}\ \bibnamefont
  {Price}},\ }\href {\doibase 10.1103/PhysRevD.5.2419} {\bibfield  {journal}
  {\bibinfo  {journal} {Phys.\ Rev.\ D}\ }\textbf {\bibinfo {volume} {5}},\
  \bibinfo {pages} {2419} (\bibinfo {year} {1972})}\BibitemShut {NoStop}%
\bibitem [{\citenamefont {Burko}\ and\ \citenamefont
  {Khanna}(2004)}]{Burko:2004jn}%
  \BibitemOpen
  \bibfield  {author} {\bibinfo {author} {\bibfnamefont {L.~M.}\ \bibnamefont
  {Burko}}\ and\ \bibinfo {author} {\bibfnamefont {G.}~\bibnamefont {Khanna}},\
  }\href {\doibase 10.1103/PhysRevD.70.044018} {\bibfield  {journal} {\bibinfo
  {journal} {Phys.\ Rev.\ D}\ }\textbf {\bibinfo {volume} {70}},\ \bibinfo
  {pages} {044018} (\bibinfo {year} {2004})},\ \Eprint
  {http://arxiv.org/abs/gr-qc/0403018} {arXiv:gr-qc/0403018} \BibitemShut
  {NoStop}%
\bibitem [{\citenamefont {Hod}(2013)}]{Hod:2013dka}%
  \BibitemOpen
  \bibfield  {author} {\bibinfo {author} {\bibfnamefont {S.}~\bibnamefont
  {Hod}},\ }\href {\doibase 10.1088/0264-9381/30/23/237002} {\bibfield
  {journal} {\bibinfo  {journal} {Classical Quantum Gravity}\ }\textbf
  {\bibinfo {volume} {30}},\ \bibinfo {pages} {237002} (\bibinfo {year}
  {2013})}\BibitemShut {NoStop}%
\bibitem [{\citenamefont {Decanini}\ \emph {et~al.}()\citenamefont {Decanini},
  \citenamefont {Folacci},\ and\ \citenamefont {Ould El~Hadj}}]{DFOEH1}%
  \BibitemOpen
  \bibfield  {author} {\bibinfo {author} {\bibfnamefont {Y.}~\bibnamefont
  {Decanini}}, \bibinfo {author} {\bibfnamefont {A.}~\bibnamefont {Folacci}}, \
  and\ \bibinfo {author} {\bibfnamefont {M.}~\bibnamefont {Ould El~Hadj}},\
  }\href@noop {} {\emph {\bibinfo {title} {Work in preparation}}}\BibitemShut
  {NoStop}%
\bibitem [{\citenamefont {Rosa}\ and\ \citenamefont
  {Dolan}(2012)}]{Rosa:2011my}%
  \BibitemOpen
  \bibfield  {author} {\bibinfo {author} {\bibfnamefont {J.~G.}\ \bibnamefont
  {Rosa}}\ and\ \bibinfo {author} {\bibfnamefont {S.~R.}\ \bibnamefont
  {Dolan}},\ }\href {\doibase 10.1103/PhysRevD.85.044043} {\bibfield  {journal}
  {\bibinfo  {journal} {Phys.\ Rev.\ D}\ }\textbf {\bibinfo {volume} {85}},\
  \bibinfo {pages} {044043} (\bibinfo {year} {2012})},\ \Eprint
  {http://arxiv.org/abs/1110.4494} {arXiv:1110.4494 [hep-th]} \BibitemShut
  {NoStop}%
\bibitem [{\citenamefont {Hassan}\ \emph {et~al.}(2013)\citenamefont {Hassan},
  \citenamefont {Schmidt-May},\ and\ \citenamefont {von
  Strauss}}]{Hassan:2012wr}%
  \BibitemOpen
  \bibfield  {author} {\bibinfo {author} {\bibfnamefont {S.}~\bibnamefont
  {Hassan}}, \bibinfo {author} {\bibfnamefont {A.}~\bibnamefont {Schmidt-May}},
  \ and\ \bibinfo {author} {\bibfnamefont {M.}~\bibnamefont {von Strauss}},\
  }\href {\doibase 10.1007/JHEP05(2013)086} {\bibfield  {journal} {\bibinfo
  {journal} {JHEP}\ }\textbf {\bibinfo {volume} {1305}},\ \bibinfo {pages}
  {086} (\bibinfo {year} {2013})},\ \Eprint {http://arxiv.org/abs/1208.1515}
  {arXiv:1208.1515 [hep-th]} \BibitemShut {NoStop}%
\bibitem [{\citenamefont {de~Rham}\ and\ \citenamefont
  {Gabadadze}(2010)}]{deRham:2010ik}%
  \BibitemOpen
  \bibfield  {author} {\bibinfo {author} {\bibfnamefont {C.}~\bibnamefont
  {de~Rham}}\ and\ \bibinfo {author} {\bibfnamefont {G.}~\bibnamefont
  {Gabadadze}},\ }\href {\doibase 10.1103/PhysRevD.82.044020} {\bibfield
  {journal} {\bibinfo  {journal} {Phys.\ Rev.\ D}\ }\textbf {\bibinfo {volume}
  {82}},\ \bibinfo {pages} {044020} (\bibinfo {year} {2010})},\ \Eprint
  {http://arxiv.org/abs/1007.0443} {arXiv:1007.0443 [hep-th]} \BibitemShut
  {NoStop}%
\bibitem [{\citenamefont {de~Rham}\ \emph {et~al.}(2011)\citenamefont
  {de~Rham}, \citenamefont {Gabadadze},\ and\ \citenamefont
  {Tolley}}]{deRham:2010kj}%
  \BibitemOpen
  \bibfield  {author} {\bibinfo {author} {\bibfnamefont {C.}~\bibnamefont
  {de~Rham}}, \bibinfo {author} {\bibfnamefont {G.}~\bibnamefont {Gabadadze}},
  \ and\ \bibinfo {author} {\bibfnamefont {A.~J.}\ \bibnamefont {Tolley}},\
  }\href {\doibase 10.1103/PhysRevLett.106.231101} {\bibfield  {journal}
  {\bibinfo  {journal} {Phys.\ Rev.\ Lett.}\ }\textbf {\bibinfo {volume}
  {106}},\ \bibinfo {pages} {231101} (\bibinfo {year} {2011})},\ \Eprint
  {http://arxiv.org/abs/1011.1232} {arXiv:1011.1232 [hep-th]} \BibitemShut
  {NoStop}%
\bibitem [{\citenamefont {Berti}\ and\ \citenamefont
  {Cardoso}(2006)}]{Berti:2006wq}%
  \BibitemOpen
  \bibfield  {author} {\bibinfo {author} {\bibfnamefont {E.}~\bibnamefont
  {Berti}}\ and\ \bibinfo {author} {\bibfnamefont {V.}~\bibnamefont
  {Cardoso}},\ }\href {\doibase 10.1103/PhysRevD.74.104020} {\bibfield
  {journal} {\bibinfo  {journal} {Phys.\ Rev.\ D}\ }\textbf {\bibinfo {volume}
  {74}},\ \bibinfo {pages} {104020} (\bibinfo {year} {2006})},\ \Eprint
  {http://arxiv.org/abs/gr-qc/0605118} {arXiv:gr-qc/0605118} \BibitemShut
  {NoStop}%
\bibitem [{\citenamefont {Leaver}(1986)}]{Leaver:1986gd}%
  \BibitemOpen
  \bibfield  {author} {\bibinfo {author} {\bibfnamefont {E.~W.}\ \bibnamefont
  {Leaver}},\ }\href {\doibase 10.1103/PhysRevD.34.384} {\bibfield  {journal}
  {\bibinfo  {journal} {Phys.\ Rev.\ D}\ }\textbf {\bibinfo {volume} {34}},\
  \bibinfo {pages} {384} (\bibinfo {year} {1986})}\BibitemShut {NoStop}%
\bibitem [{\citenamefont {Andersson}(1997)}]{Andersson:1996cm}%
  \BibitemOpen
  \bibfield  {author} {\bibinfo {author} {\bibfnamefont {N.}~\bibnamefont
  {Andersson}},\ }\href {\doibase 10.1103/PhysRevD.55.468} {\bibfield
  {journal} {\bibinfo  {journal} {Phys.\ Rev.\ D}\ }\textbf {\bibinfo {volume}
  {55}},\ \bibinfo {pages} {468} (\bibinfo {year} {1997})},\ \Eprint
  {http://arxiv.org/abs/gr-qc/9607064} {arXiv:gr-qc/9607064} \BibitemShut
  {NoStop}%
\bibitem [{\citenamefont {Andersson}\ and\ \citenamefont
  {Glampedakis}(2000)}]{Andersson:1999wj}%
  \BibitemOpen
  \bibfield  {author} {\bibinfo {author} {\bibfnamefont {N.}~\bibnamefont
  {Andersson}}\ and\ \bibinfo {author} {\bibfnamefont {K.}~\bibnamefont
  {Glampedakis}},\ }\href {\doibase 10.1103/PhysRevLett.84.4537} {\bibfield
  {journal} {\bibinfo  {journal} {Phys.\ Rev.\ Lett.}\ }\textbf {\bibinfo
  {volume} {84}},\ \bibinfo {pages} {4537} (\bibinfo {year} {2000})},\ \Eprint
  {http://arxiv.org/abs/gr-qc/9909050} {arXiv:gr-qc/9909050} \BibitemShut
  {NoStop}%
\bibitem [{\citenamefont {Glampedakis}\ and\ \citenamefont
  {Andersson}(2001)}]{Glampedakis:2001js}%
  \BibitemOpen
  \bibfield  {author} {\bibinfo {author} {\bibfnamefont {K.}~\bibnamefont
  {Glampedakis}}\ and\ \bibinfo {author} {\bibfnamefont {N.}~\bibnamefont
  {Andersson}},\ }\href {\doibase 10.1103/PhysRevD.64.104021} {\bibfield
  {journal} {\bibinfo  {journal} {Phys.\ Rev.\ D}\ }\textbf {\bibinfo {volume}
  {64}},\ \bibinfo {pages} {104021} (\bibinfo {year} {2001})},\ \Eprint
  {http://arxiv.org/abs/gr-qc/0103054} {arXiv:gr-qc/0103054} \BibitemShut
  {NoStop}%
\bibitem [{\citenamefont {Finn}\ and\ \citenamefont
  {Sutton}(2002)}]{Finn:2001qi}%
  \BibitemOpen
  \bibfield  {author} {\bibinfo {author} {\bibfnamefont {L.~S.}\ \bibnamefont
  {Finn}}\ and\ \bibinfo {author} {\bibfnamefont {P.~J.}\ \bibnamefont
  {Sutton}},\ }\href {\doibase 10.1103/PhysRevD.65.044022} {\bibfield
  {journal} {\bibinfo  {journal} {Phys.\ Rev.\ D}\ }\textbf {\bibinfo {volume}
  {65}},\ \bibinfo {pages} {044022} (\bibinfo {year} {2002})},\ \Eprint
  {http://arxiv.org/abs/gr-qc/0109049} {arXiv:gr-qc/0109049} \BibitemShut
  {NoStop}%
\end{thebibliography}%

\end{document}